# The Digital Humanities Unveiled: Perceptions Held by Art Historians and Computer Scientists about Computer Vision Technology

Emily L. Spratt
Department of Art and Archaeology, Princeton University

Ahmed Elgammal
Department of Computer Science, Rutgers University

Although computer scientists are generally familiar with the achievements of computer vision technology in art history, these accomplishments are little known and often misunderstood by scholars in the humanities.[1] To clarify the parameters of this seeming disjuncture, I and my collaborator, Ahmed Elgammal, in the recent article "Computational Beauty: Aesthetic Judgment at the Intersection of Art and Science," addressed the concerns that one example of the digitization of the humanities poses on social, philosophical, and practical levels.[2] In support of our assessment of the perceptions held by computer scientists and art historians about the use of computer vision technology to examine art in this study, we based our interpretations on two surveys that were distributed in August 2014 to the

---

I am very grateful for the editorial feedback and invaluable assistance that Denis Cummings and Sharon Herson have provided me on this project. I also thank Ahmed Elgammal for reviewing the statistical data with me and for his collaboration on this project.

[1] This paper was self-published on October 20, 2014, for the purposes of making this research related to computer vision technology in the humanities accessible to a wider audience. All rights to this paper are reserved by the creators of The Digital Humanities Project: Aesthetics at the Intersection of Art and Science website.

[2] Emily L. Spratt and Ahmed Elgammal, "Computational Beauty: Aesthetic Judgment at the Intersection of Art and Science," *Computer Vision: ECCV 2014: 13th European Conference, Zurich, Switzerland, September 6–12, 2014, Proceedings* (2014). This paper may be accessed on our website's home page: https://sites.google.com/site/digitalhumanitiessurvey/home.

departments of computer science and art history at Princeton University, Rutgers University, Cornell University, New York University, and the University of California at Los Angeles. In the present paper, the development of these surveys and their results are discussed in the context of the major conclusions (reviewed in brief) of the aforementioned publication.[3]

**The Genesis of an Unexpected Collaboration**

The initiation of this digital humanities project stems from an unexpected encounter that occurred at one of the Media and Modernity Program conferences held at the Princeton School of Architecture in 2013. It was the extreme surprise—and even discomfort—experienced by a Renaissance-Baroque and Byzantine art historian in being asked a few simple questions by a computer scientist that launched this investigation of one intersection of the arts and sciences.

I was asked: Have you ever heard of computer vision? How can computer scientists further aid art historians in their analysis of art? And, now that we have been able to analyze paintings and can group them accordingly, what would you as an art historian like to do with this data? My immediate response was essentially sheer astonishment; not only had I never heard of this type of technology, I was shocked to learn of its capabilities, which include "the automatic classification of art to identify the hand of an artist, the ability to classify paintings according to style and to distinguish stylistically similar images of paintings, the quantification of the degree of artistic similarity found between paintings, and the capability to predict a painting's date of production."[4] I was intrigued, skeptical, and even territorial—a response I soon learned was shared by many of my colleagues in art history. It was this perfect storm of conditions that launched a new research project and the dawn of a collaboration with my interrogator, Ahmed Elgammal, associate professor of computer science at Rutgers University.

---

[3] For statistical information on both surveys, see the survey results section of our website: https://sites.google.com/site/digitalhumanitiessurvey/surveyresults.

Since this fortuitous encounter, my collaborator and I have not only addressed the answers to these questions, but have also examined the disunion between our fields and the associated unease that this intersection of art and science heralds. We set out to discover whether my reaction to the discovery of what computer vision is accomplishing was an accurate indication of most art historians' perceptions of this new technology and what computer scientists thought about its application in art history. We wanted to investigate why there is an apparent disjuncture between the fields and to determine how they could be fruitfully brought together.

**Surveying Art Historians and Computer Scientists on the Use of Computer Vision Technology**

We therefore designed twenty-one questions for art historians and sixteen for computer scientists that were intended to shed light on field members' knowledge of the capabilities and applications of computer vision technology, attitudes and perceptions about the use of it, and reactions to the meaning of this type of digitization of the humanities. (The statistical results of these surveys may be examined in the "survey results" section of our website.) Although these surveys were intended to gauge perceptions of both art historians and computer scientists, the focus of our study was on the receptivity of art historians to computer vision technology; nevertheless, it was also necessary to measure computer scientists' ideas on the subject as a basic point of comparison. In both surveys, questions were intentionally structured to overlap one another in an attempt to capture more nuanced opinions of the survey participants rather than reproducing vague generalities, acknowledging that art historians in particular would be highly sensitive to the use of language in the given questions. On account of the lack of clarity regarding the use of computer science in the humanities, we also provided space for respondents to elaborate on their interpretation of each question and encouraged general remarks at the end of the survey. In order to protect the anonymity of the participants, individual comments have been summarized or appropriately selected in this paper rather than published verbatim with the statistical results of the survey. We gave the survey participants only four response choices, two for approval and two for disapproval, leaving no option for a neutral, neither agree nor disagree, position, although respondents were allowed to skip questions. In this sense, we pushed participants to take sides on the subject and thus inadvertently fostered the generation of a plethora of written comments to the questions at hand. In sum, we received seventy-five largely passionate responses, fifty-nine from art historians and sixteen from computer scientists.

---
[4] See "Computational Beauty," sections 1 and 2; for the quote, p. 3.

      While there is little surprise that the fields of computer science and art history have different perspectives, overall we found that there is much agreement on some components of the digitization of the humanities. For example, when

members from each field were asked about their receptivity to collaboration in either computer science or art history, the results revealed that the majority of members in each group were positive about working together. Nonetheless, the question "Do you think that the implementation of AI technology in the humanities signals the beginning of a positive paradigm shift in academia?" showed that more than half the art historians disapproved of this notion, whereas the overwhelming majority of the computer scientists predictably believed that it did.[5]

**The Arts and Sciences: A History of Collaboration Endures**

It is likely that the positive response to the idea of collaboration between the fields is connected to the long tradition of art history's partnership with science.[6] Although the birth of art history is usually associated with the Renaissance and Giorgio Vasari's writing of *Le vite de' più eccellenti architetti, pittori, et scultori*,[7] first published in 1550, how we define the origins of the discipline differs greatly according to the artistic tradition being considered, thus nuancing any standardization of what is meant by art historical analysis. In the West, Greek philosophers such as Plato and Aristotle could even be credited with engaging in an early form of art history, commenting at length on the faculties of observation gained through sight and the physical drives associated with seeing.[8] Renaissance humanists such as Marsilio Ficino (1433–1499) also commented on ancient philosophies of seeing and the mechanisms of human perception.[9] Together their commentaries on the faculties of observation gained through sight are suggestive of the most fundamental questions of the field: how do we see and what constitutes seeing? These basic concepts, which are at the core of aesthetic theory, were of concern to the eighteenth-century philosophers whose writings clarified the parameters of an ill-defined field. To give one example, Carl Linnaeus (1707–1778), the founding father of modern taxonomy, may be credited with establishing the foundations for the classification of artifacts in museums.[10]

---

[5] See question 19 in the art history survey and question 14 in the computer science survey.
[6] This subject has been discussed in greater depth in "Computational Beauty."
[7] Giorgio Vasari, *The Lives of the Artists*, trans. Julia Conaway Bondanella and Peter Bondanella (Oxford: Oxford University Press, 1991).
[8] Aristotle, *Poetics*, trans. Anthony Kenny (Oxford: Oxford University Press, 2013).

While the period from the sixteenth century to the end of the nineteenth witnessed many methodological developments in the history of art, these contributions were largely philosophical and less emulative of the direction taken by Linnaeus. It was not until the nineteenth century that the first thorough reassessment of the techniques and standards of connoisseurship established by Vasari was conducted by Italian art historian Giovanni Morelli (1816–1891).[11] Morelli critiqued a fracture within the field between a theory-based, "armchair" art history versus an object-centered, "laboratory" approach. His innovation was to focus on methods of connoisseurship that privileged direct engagement with a work of art that allowed for a very precise type of visual investigation. Not surprisingly, his theories found immediate application in the world of connoisseurs, conservators, and museum associates. In the same vein, these types of materialist inquiries opened theoretical ground for philosophical consideration of the history of art measured through the development of form itself, devoid of its socio-historical constraints. In essence, there was a divide between theoretical art historians and connoisseurs. More than a hundred years later, there remains no consensus on how much connoisseurship should inform theory and vice versa. These differences in opinion about how art history should be practiced were, in fact, articulated by the majority of the art historian survey commentators. By extension, this directly influenced how each respondent interpreted and valued the developments in computer vision technology.[12]

---

[9] Marsilio Ficino, "'Proem' to Commentary on Plotinus," in *Opera Omnia*, 2 vols. (Basel, 1576).
[10] For more on this subject, see Paula Findlen, *Possessing Nature: Museums, Collecting, and Scientific Culture in Early Modern Italy* (Berkeley: University of California Press, 1994).
[11] See Giovanni Morelli, *Kunstkritische Studien über italienische Malerei*, Band 1, *Die Galerien Borghese und Doria Pamfilj in Rom* (Leipzig: F. A. Brockhaus, 1890); Giovanni Morelli, *Italian Painters: Critical Studies of Their Works*, trans. Constance Jocelyn Ffoulkes (London: John Murray, 1892).

**Artificial Intelligence: A Different Type of Scientific Application in Art History**

While art history has long been engaged with the sciences, the developments in computer science today offer something notably different from empirical inspiration. Since the pioneering research on two-dimensional imaging for statistical pattern recognition brought the computer from a typewriting calculator to an image-processing machine that can see, it is clear that computer science offers something radically new to the humanities. At this point in time, computer science has transcended mere physical and computational assessments of art, and draws upon multiple forms of visual analysis that together make complex interpretations of a given object or group of objects.[13] Our survey indicated that art historians were mostly unaware of these advances and were only familiar with the most basic applications of computer science in their field. Given that computers can now perform some of the core processes of aesthetic interpretation that an art historian would make, the first question is not how does art history use these new methods, but what does this mean in terms of aesthetic interpretation itself?

---

[12] These comments were made throughout the survey.

      Since the very nature of our ability to aesthetically comprehend and judge beauty is the determining factor in what most people would describe as distinguishing us from machines, this type of computer science threatens our own conceptions of human identity. Not surprisingly, art historians responded territorially to the idea of computer vision technology applications. For example, one commentator explicitly stated the extreme position: "Why would I entrust my life's work to a computer just so I could be made irrelevant."[14] When art historians were asked generally whether they would be open to the use of computers to make aesthetic judgments about art, the overwhelming majority responded that they would not. While it is important to recognize these anxieties, we would like to first propose that understanding some of the philosophical origins of how we in the humanities have come to regard aesthetic judgment may offer an explanation as to why it is that persons not trained in computer science perceive these developments with suspicion. Second, computer science has presented the humanities a challenge: is intangible, or sensory, knowledge really intangible if a computer can perform

---

[13] See section 2 of "Computational Beauty" for a comprehensive assessment of these technological developments.

[14] See question 6 in the art history survey.

processes that manifest the same results that a human would produce? In this paper, these complex debates are presented in brief.

**Philosophy and Aesthetics**

The concept of sensory knowledge derives from a long tradition, although it was not until the eighteenth century that this type of knowing began to be perceived in a positive light. Predominantly on account of Alexander Gottlieb Baumgarten's *Aesthetica*, published in 1750, the notion that there was a type of knowledge distinct from that of logic or reason gained acceptance.[15] He termed this knowledge an analogue of reason, which had its own perfection distinct from logic. This theory led to the proposition that there should be two kinds of corresponding sciences of knowledge: logic and aesthetics. Baumgarten's philosophy provided the foundation for Immanuel Kant's theories on aesthetics and the immediate background for the *Critique of Judgment* published in 1790.[16]

The key to Kant's discourse was his rooting of the condition of aesthetic discernment in a subjective, non-logical process. Indeed, the philosophy of aesthetics from Baumgarten to Gilles Deleuze, not necessarily including aesthetic conceptions in Hegelian philosophy, places aesthetic comprehension in the realm of subjectivity.[17] Kant articulated the conditions of this type of reasoning in the

---

[15] Alexander Gottlieb Baumgarten, *Aesthetica* (Hildesheim: Georg Olms Publishing House, 1961).
[16] Immanuel Kant, *Critique of Aesthetic Judgment*, trans. James Creed Meredith (Oxford: Clarendon Press, 1911).
[17] Hegel regarded art as "a secondary or surface phenomenon . . . thus harking back to pre-Baumgarten and pre-Kantian ideology which privileged the ideal or Thought

*Critique of Judgment,* locating aesthetic understanding in moral philosophy and the principles of universality.[18] In part 1 of the *Critique,* Kant explains that "[t]he judgment of taste . . . is not a cognitive judgment, and so not logical, but is aesthetic—which means that it is one whose determining ground cannot be other than subjective."[19]

Despite the focus on the subjectivity of aesthetic interpretation through individual judgment, Kant goes on to clarify that the judgment of taste is also universal. He suggests that beauty is linked to the infinite quality of the human imagination yet grounded in the finiteness of being. In this sense, the universality of taste is also related to a type of collective consciousness that stems from God's universal creation. Kant further relates aesthetics and ethics, positing that beautiful objects inspire sensations like those produced in the mental state of moral judgment; thus, genius and taste could be related to the moral character of an artist or viewer.[20] How moral values can raise or lower the aesthetic value of art is, indeed, a subject of philosophical scrutiny, if not controversy, to this day. Although there is no consensus in art history as to whether there is a universal sense of beauty given the diversity of cultural values, this question has not been of great concern in the field for quite some time and is more related to issues of connoisseurship.

---

by devalorizing visual knowledge"; see Donald Preziosi, *The Art of Art History: A Critical Anthology* (Oxford and New York: Oxford University Press, 1998).
[18] Kant, *Critique of Aesthetic Judgment*.
[19] Ibid., 41.
[20] These interpretations were facilitated by Donald Preziosi; see Preziosi, *The Art of Art History*, 66–67.

Instead, art history, the discipline, has a tradition of intellectual borrowings for its theories and methodologies, incorporating the perspectives of diverse fields in the humanities. Due to the field's inherent flexibility, critical interpretations rarely produce a singular analysis of art. Although parallel interpretations of a given object are implicitly understood to exist, a Kantian aesthetic background still pervades many theories of art. Because art historians generally assume that visual processes are wholly subjective, how can they accept the idea that computer vision technology offers aesthetic interpretations when such a notion betrays their philosophical understanding of that mental process?

Indeed, the overwhelming majority of art historians that responded to the survey thought it was not possible for a computer to analyze art in terms of beauty, style, dating, and relative influence in the development of art through history.[21] One commentator went so far as to say that the idea of a computer assessing beauty sounded "dangerous since 'beauty' means different things to different people and cultures."[22] Another commentator thought that a computer could facilitate this type of work as an aid to the art historian but not as a stand-alone replacement. People emphasized their belief that computers can aid in conservation analysis of a painting but not in terms of dating, beauty, or relative comparisons to other works of art. Overall, art historians had very different ideas about what a computer was capable of doing, but the majority of commentators were clear that interpreting the beauty of something was a subjective judgment that only a human could perform.

---

[21] See question 3 in the art history survey.
[22] Ibid.

Now that computers are able to perform these types of tasks, however, the field needs to revisit what this type of aesthetic interpretation means.

**Aesthetic Judgment Readaddressed**

The machine's ability to make an aesthetic judgment about a painting and then compare it stylistically to other paintings demonstrates that logic and objectivity—not intangible subjectivity—*are* at work in the complicated algorithms that compose the artificial intelligence system. These processes are all clearly imitative and objective at the point of the computer program training period; once the machine reaches the automaton level, however, the question of subjectivity reenters. In this sense, are computer programmers like blind watchmakers, to use Richard Dawkins's famous metaphor of the evolution of the universe and the free will debate?[23] Are computers comparable to humans with genetic codes that predetermine outcomes, which are then shaped by the environment?

    Fortunately, a new area of research, *neuroaesthetics*, has begun to explore the biological processes that underlie visual interpretation, and similarities of neural network communication with computer programming models have been noted.[24] In light of these discoveries, a number of questions emerge: if we are able to create artificial intelligence that performs types of reasoning that we have long considered

---

[23] Richard Dawkins, *The Blind Watchmaker: Why the Evidence of Evolution Reveals a Universe Without Design* (New York: W. W. Norton & Company, 1996).
[24] See for instance, John Onians, *Neuroarthistory: From Aristotle and Pliny to Baxandall and Zeki* (New Haven: Yale University Press, 2007); David Freedberg, *The Power of Images: Studies in the History and Theory of Response* (Chicago: University of Chicago Press, 1989). See also New York University's Center for Neural Science and the Visual Neuroscience Laboratory: http://www.cns.nyu.edu/

subjective, we are more machine-like than we admit, machines have more human potential than we estimate, or these processes are, in fact, tangibly measurable and objectively determined.[25] In essence, the debate moves to the question of determinism and free will. While most people would agree that a computer, even one that has reached automaton status and has the ability to learn from its environment, is not free, we are less willing to concede the notion of human freedom when we, too, are ultimately bound by our genes and environment.

If our understanding of aesthetic judgment is still tied to the eighteenth- and nineteenth-century philosophical tradition, at the least we need to better interpret how these subjective processes work, determine if they really are subjective, and integrate new scientific developments, such as those in neurobiology and computer science, into our conceptions of how *knowledge* is produced. It is a paradox that developments in computer science could push the humanities to reevaluate its most basic premises even if they are not currently the trendy questions for an art historian to be asking: for art history, this is how we determine that something is beautiful and/or important, and how objects are interrelated. Have the advances in science not provided a platform in which we can begin to understand cognition, as it is applied to aesthetics, in a radically different way than eighteenth-century philosophers conceived these processes? We easily discredit the idea of humors as ruling temperaments of the body but know that Kant considered them viable and one of them as an indication of the absence of temperament.[26] We still read Kant for his interpretations of physical and psychological states, yet not on his theory of the phlegmatic humor.

---

[25] Frederick Turner, in his *Beauty: The Value of Values* (Charlottesville: University Press of Virginia, 1991), has made the extreme suggestion that aesthetic interpretation is an entirely objective process.

Not surprisingly, in the survey to art historians, their response to the general question "Would you be open to the use of computers to make aesthetic judgments about art?" was very negative.[27] A number of commentators explicitly stated that aesthetic judgments are inherently subjective. The question, however, remains open that if style, subject matter, dating, and even beauty are not part of an aesthetic judgment, then how are we defining aesthetic judgment? Considering the positive responses to questions about particular components of an aesthetic judgment, we conclude that art historians are willing to consider the possibility of a computer making specific interpretations in the realm of aesthetics but that the concept of these tasks being performed in unison or on the level of automated computer analysis remains mostly unimaginable. This interpretation would support our survey results that art historians were open to the idea of using a computer to make analyses of paintings on the level of stylistic interpretation but only when specific categories were clearly highlighted and articulated. Indeed, art historians remained suspicious about the use of computer vision technology when presented with the general concept of a computer performing aesthetic judgments. This finding thus suggests how members of the humanities' perceptions about digitization could be altered: on the level of specifics.

---

[26] Robert B. Louden, *Kant's Impure Ethics: From Rational Beings to Human Beings* (New York: Oxford University Press, 2000), 79.
[27] See question 8 in the art history survey.

**Conclusions on the Implementation of Computer Vision Technology in Art History**

In conclusion, we would like to underscore the current concerns that the immediate application of this research poses for art historians. We have thus highlighted three main issues that demand further attention: the use of language between fields to describe concepts; the problem in the lack of uniformity in the interpretation of art; and the separate trajectories within computer science and art history regarding aesthetic interpretation.

First, there is discomfort in the globalizing language that computer scientists use to describe their research. The use of language to discuss digitizing the humanities has had a large impact on receptivity to the notion of change in field practices. Rather than make claims about a computer's ability to analyze art at large, specificity as to what can be analyzed and what has been analyzed would assuage anxieties about the ontological nature of the human versus the machine.[28] For example, when art historians were questioned about the use of computers in areas traditionally associated with connoisseurship and conservation that are well known in the field, the response was overwhelmingly positive. Given that this response was one of our hypotheses in writing the survey, we also asked two questions regarding drone technology to see if we could obtain a clear statistical reading of this probable phenomenon. We therefore asked one general question about the development of drones and the other in regard to its application to archaeology to determine whether knowledge of the specific use of technology was less daunting than the general idea of it. Indeed, just over half the art history respondents did not support the development of drone technology, whereas more than 85 percent approved of its use specifically for archaeological discovery. In the same vein, both art historians and computer scientists thought that computers should be used as a tool rather than as a total replacement for specialists. Interestingly, neither art historians nor computer scientists thought it likely that a computer could take away an art historian's job, despite the appearance in the press of sensationalizing article titles.[29]

---

[28] Emily L. Spratt, "Man versus Machine: Aesthetic Judgment in the Age of Artificial Intelligence" (presentation, Theoretical Archaeology Annual Group Meeting, University of Illinois at Urbana-Champaign, May 23–25, 2014).

      We would therefore like to propose that computer scientists collaborate with art historians on specific projects. The ability to compute perspective coherence, lighting and shading strategies, brushstroke styles, and semantic points of similarity could, for example, aid the analysis of a large group of Italian drawings with unclear authorship. Similarly, the application of this technology for the identification of icon workshops that utilized the same iconographic templates in the context of Byzantine devotional images would be extremely useful if a large dataset of images from diverse collections that are not readily accessible to the public could be brought together. Recent collaborations of this nature have already been initiated

---

[29] Matthew Sparkes, "Could Computers Put Art Historians Out of Work?" *The Telegraph*, Aug. 18, 2014 (http://www.telegraph.co.uk/technology/news/11041814/Could-computers-put-art-historians-out-of-work.html).

and should continue, in hopes of motivating more collection holders to digitize their objects.[30]

The second issue concerns the way the social history of an object and the emotional engagement to art is calculated. In art history, the degree to which the context in which a work of art is produced should matter. How can a computer quantify the social history of a painting? Furthermore, if our understanding of the history of art is related to the emotional response that an object elicits, how can a computer mimic human affect? That what we have understood as subjective processes may in fact be objectively determined problematizes the argument that computers can never achieve the capacities beholden to the contemplative human mind.

While there are current limitations of the application of computer vision technology, certain periods or genres are more amenable to its current capabilities. For example, Abstract Expressionism, which is highly concerned with the role of form over content, naturally accommodates the high degree of stylistic interpretation that computer vision offers. Automatic influence detection has demonstrated the ability to detect less overt connections between artists, such as Eugene Delacroix's not so widely known influence from El Greco in terms of both color and expressiveness. While this observation highlights the remarkable subtleties of interpretation that computer vision is capable of generating, this type of analysis is of less use to an art historian than a more specific study, such as what an analysis of Kazimir Malevich's fairly uniform-appearing Suprematist paintings might reveal in regard to style.

---

[30] See Jia Li, Lei Yao, Ella Hendriks, and James Z. Wang, "Rhythmic Brushstrokes Distinguish van Gogh from His Contemporaries: Findings via Automated Brushstroke Extraction," *IEEE Transactions on Pattern Analysis and Machine Intelligence* 34, no. 6 (2012): 1159–76; J. M. Hughes, Dong Mao, D. N. Rockmore, Yang Wang, and Qiang Wu, "Empirical Mode Decomposition Analysis for Visual Stylometry," *IEEE Transactions on Pattern Analysis and Machine Intelligence* 34, no. 11 (2012): 2147–57.

But to what degree can we ascribe the detection of influence or artistic merit to a machine when it was the computer scientists who wrote the programming that associated certain visual components with particular markers of identity? At what point in the process of "training" the program to make its own judgments does the machine develop autonomy, if ever? If computer scientists can be charged with owning the responsibility of artistic interpretation at the level of programming input, why wouldn't art historians be involved at this level of the research?

David Hume philosophized that ``beauty is no quality in things themselves: it exists merely in the mind which contemplates them; and each mind perceives a different beauty."[31] If the interpretation of art lies in the eyes of the beholder and is thus a subjectively determined process that is associated with feeling, how can we understand the development of autonomous aesthetic evaluation from a computer? Computer scientists and art historians both agreed that the humanities should be more digitized, but before art historians are willing to believe it is possible to analyze art with a computer in terms of beauty, style, dating, and relative influence in the development of art through history, we must revisit aesthetic judgment.

Fortunately, at this particular intersection of the arts and sciences, future entanglement between computer science and art history is promising. Let us not underestimate art historians' interest in computer science—commentators of the survey repeatedly stated that they wished they knew more about the movement of artificial intelligence into the humanities—or the attention to art history by computer scientists, of which half the group surveyed indicated that they were interested to get direct feedback from art historians as they developed new applications for computer vision technology.[32]

---

[31] David Hume, *Of the Standard of Taste and Other Essays*, ed. John W. Lenz (Indianapolis: Bobbs-Merrill, 1965).

---

[32] These remarks were made in the general comments section of both surveys and in question 13 in the computer science survey.